\documentclass[epj]{svjour}

\usepackage{graphicx}
\usepackage{amsmath}
\usepackage{epstopdf}

\begin{document}

\title{A model for the onset of oscillations near the stopping angle in an inclined granular flow}

\author{D. Tan\inst{1} \and P. Richard\inst{2} \and J. T. Jenkins\inst{1}}

\institute{                    
  \inst{1} Field of Theoretical and Applied Mechanics, Cornell University, Ithaca, New York 14853, USA \\
  \inst{2} Institut de Physique de Rennes, UMR CNRS 6251, University of Rennes 1, 35042 Rennes, France \\
}

\date{Received: date / Revised version: date}

\abstract{
We propose an explanation for the onset of oscillations seen in numerical simulations of dense, inclined flows of inelastic, frictional spheres. It is based on a phase transition between disordered and ordered collisional states that may be interrupted by the formation of force chains. Low frequency oscillations between ordered and disordered states take place over weakly bumpy bases; higher-frequency oscillations over strongly bumpy bases involve the formation of particle chains that extend to the base and interrupt the phase change. The predicted frequency and amplitude of the oscillations induced by the unstable part of the equation of state are similar to those seen in the simulations and they depend upon the contact stiffness in the same way. Such oscillations could be the source of sound produced by flowing sand.
}

\maketitle

\section{Introduction \label{intro}}
Inclined granular flows are a common subject of experiments and numerical simulations. Oscillations in such flows are of particular interest because they could be the source of sound produced during spontaneous avalanches.

A periodic three-dimensional numerical simulation by Silbert \cite{silbert1} of a dense, inclined granular flow exhibited significant oscillations in kinetic energy as the inclination was decreased to within two degrees of the stopping angle. These high-frequency oscillations were also present in both the normal velocity autocorrelation and inter-particle contact force time-correlation functions. Finally, the coordination number was seen to vary periodically between one and four, which indicates particles forming and breaking contacts, implying the formation of particle chains. Indeed, Silbert's fig.\ 5 \cite{silbert1} showed the cyclic appearance of `force chains' in the lower part of the numerical simulation: at a peak in coordination number, there is a dense network of contacts in the lower half of the flow; and when the coordination number is at a minimum, the network of contacts is significantly less dense. This indicates that force chains are formed (causing the dense network) and broken (the less dense network) over time. Another set of similar simulations by Silbert, et al.\ \cite{silbert2}, but which involved a weakly bumpy base, exhibited oscillations at lower frequencies between the random and ordered states without any chains forming.

Mills and Chevoir \cite{millschevoir} attributed the high frequency oscillations reported by Silbert \cite{silbert1} to a decrease in stress ratio with inertia parameter near the jamming transition, where enduring contact forces dominate instantaneous collisions. Oscillations then take place between the densest possible state and a less dense state, both involving force chains that span the system. Here, we consider the oscillations to be associated with the first-order phase transition in a hard sphere gas that takes place between collisional states \cite{alder,rintoul} and interpret the appearance of the oscillations as resulting from the volume fraction being forced onto the unstable branch of the equation of state. This phase change takes place between a random collisional state in which the steady-state volume fraction is lower, and an ordered collisional state, in which the steady-state volume fraction is higher. Decreasing the inclination angle in Silbert's simulations is equivalent to increasing the steady-state solid volume fraction and, beyond a certain value, oscillations of significant amplitude, associated with the phase change, are initiated. When the base is strongly bumpy, this phase change may be interrupted by the formation of force chains. Hence, the ordered collisional state does not appear in Silbert's fig.\ 5 \cite{silbert1}.

\begin{figure}   
	\begin{center}
	\includegraphics[width=2.5cm]{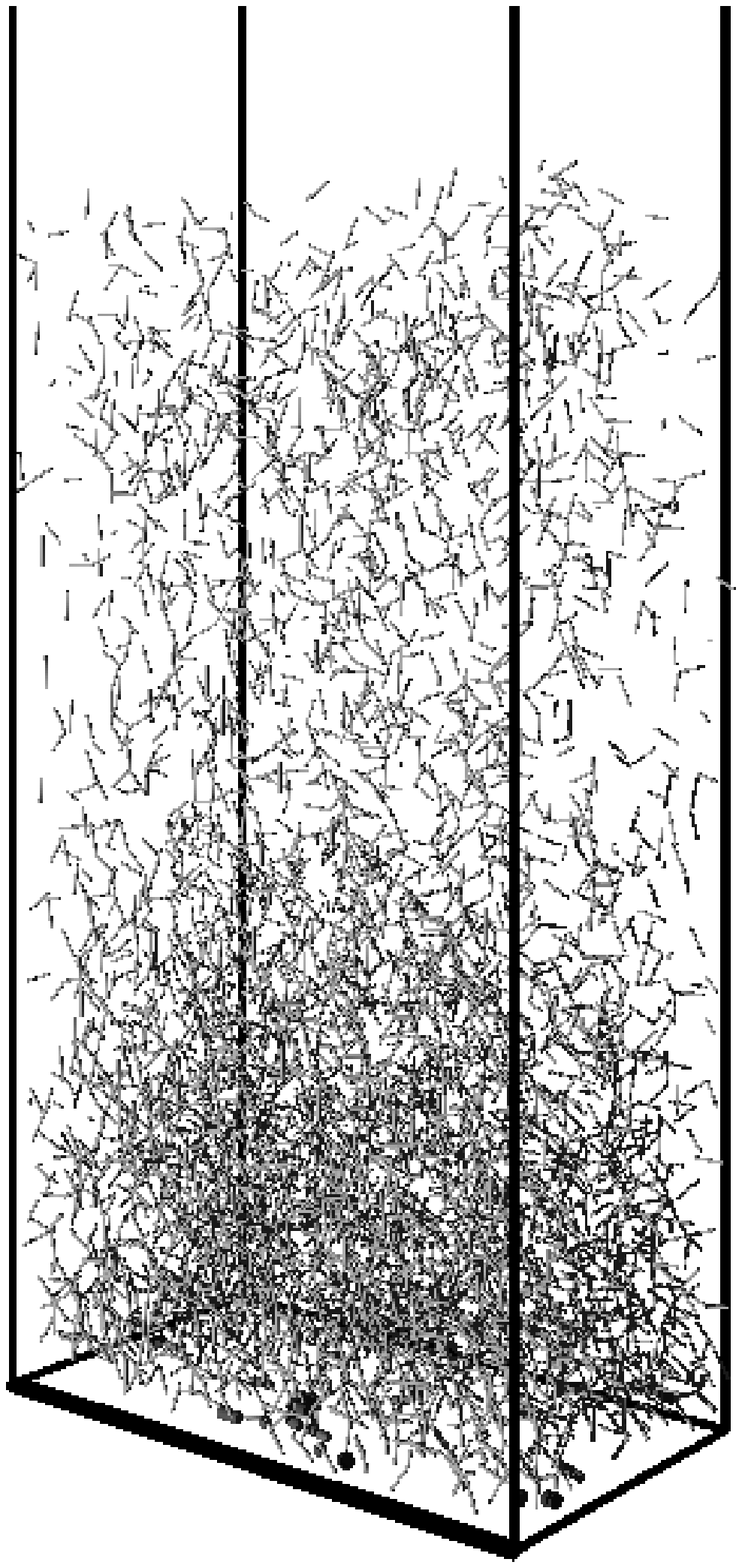}
	\hspace{1cm}
	\includegraphics[width=2.5cm]{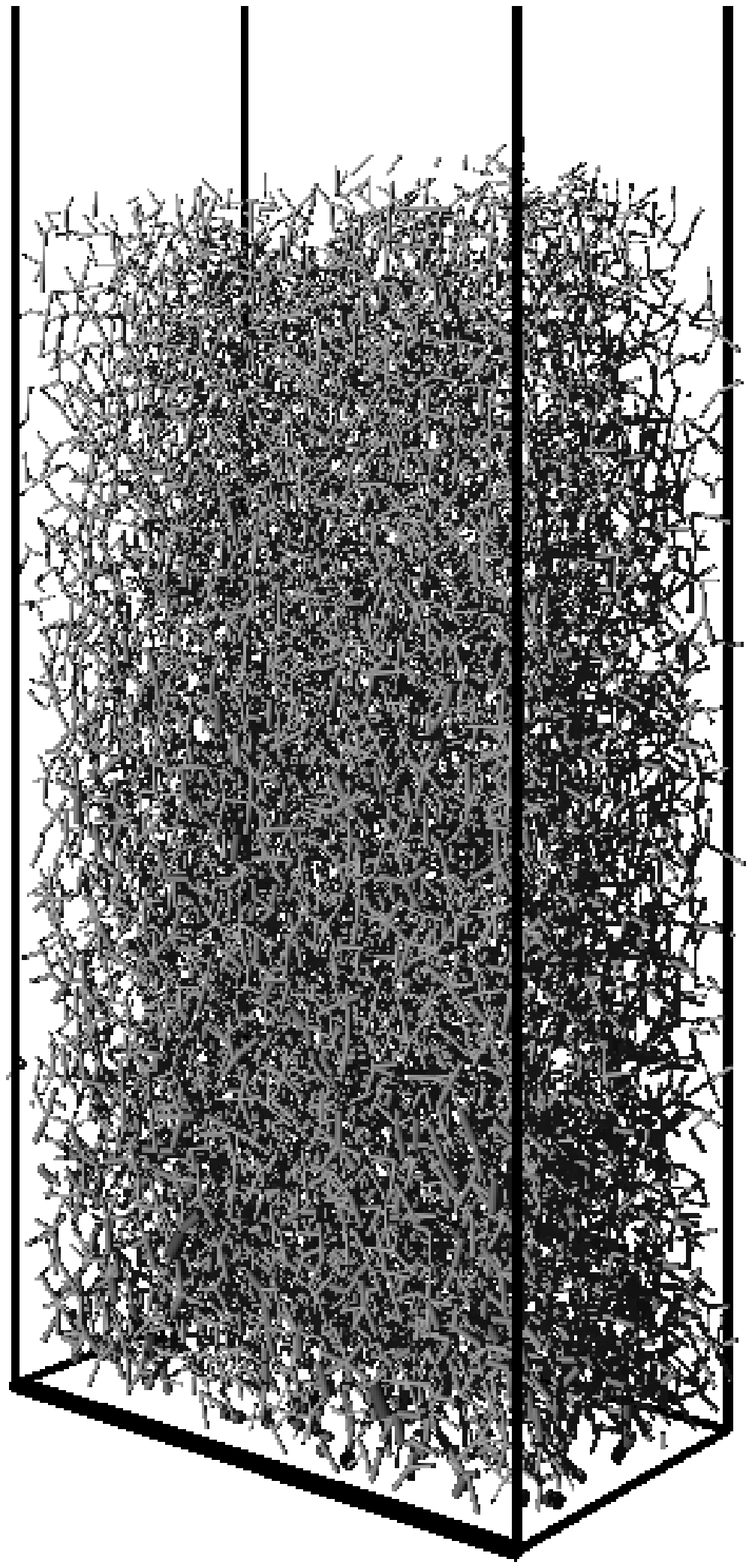}
	\caption{Snapshots of force chains present during oscillations observed in a numerical simulation of gravity-driven chute flow at inclination $19.5^{\circ}$ (see animated gif, 1.74 MB in supplementary material).  The simulation was similar to those described in \cite{silbert1} and \cite{patrick}.  \label{patrick-forcechains1}}
	\end{center}
\end{figure}

Figure \ref{patrick-forcechains1} shows two phases in the oscillation over a bumpy base. The first involves small clusters or chains of overlapping particles; these are of limited extent and do not span the flow. The second exhibits chains of greater length that extend to the base. The particle chains that extend to the base of the flow provide a stiffer resistance to changes in volume fraction than the compression of the dense, collisional gas. When the base is weakly bumpy, but frictional, the formation of particle chains is suppressed and the oscillations take place between ordered and disordered states of the dense gas of colliding particles.

\begin{figure}
	\begin{center}
	\includegraphics[width=8cm]{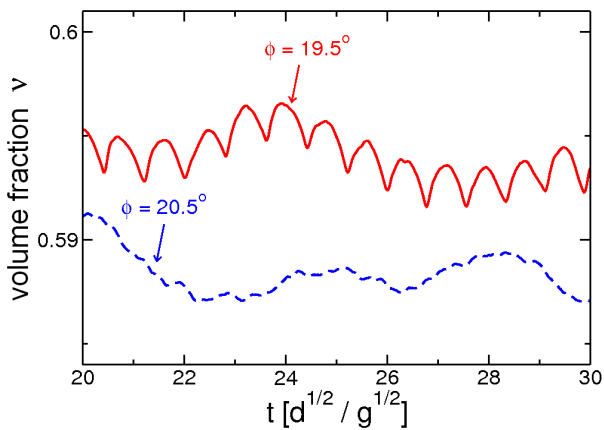}
	\caption{(Color online) Volume fraction against time at depth 33d$<z<$35d (measured from the base) for inclinations $19.5^{\circ}$ and $20.5^{\circ}$, in numerical simulations of a gravity-driven chute flow similar to those described in \cite{silbert1} and \cite{patrick}.	\label{19o5-20o5deg}}
	\end{center}
\end{figure}

Figure \ref{19o5-20o5deg} shows the solid volume fraction at the same depth in two gravity-driven chute flows of different inclination (19.5$^\circ$ and 20.5$^\circ$), from numerical simulations similar to those conducted by Silbert \cite{silbert1}. The bumpy base was formed by randomly gluing particles onto a flat base, resulting in a boundary that was more bumpy than that of Silbert, et al.\ \cite{silbert2}, but probably less bumpy than the free surface of a granular pile of Silbert \cite{silbert1}. The flow was 40 particle diameters deep, and the particles were given a stiffness of $2\times 10^5$ $mg/d$. As expected, the volume fraction was higher for the lower inclination. However oscillations of significant amplitude appeared throughout the flow depth, which were not present in the higher inclination flow.

\section{Homogeneous Phase Transition \label{Homogeneous}}
Both the volume fraction and the oscillations of interest are very similar throughout the flow, so the assumption of a homogeneous profile as a first approximation is appropriate.

\begin{figure}
	\begin{center}
	\includegraphics[height=5.5cm]{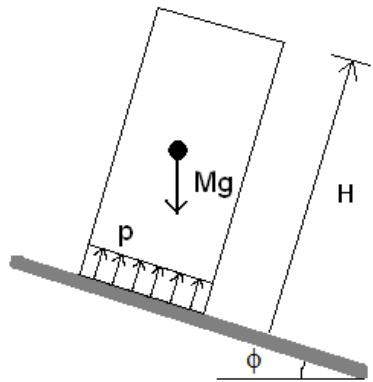}
	\caption{Schematic of the homogeneous phase transition mode. $\phi$ is the inclination angle, $H$ is the height of the flow, $M$ its mass and $p$ the pressure. \label{HomPhaseTrans}}
	\end{center}
\end{figure}

Here, we model a portion of a dense inclined gravity-driven chute flow of height $H=H_0+H_1(t)$, base cross-sectional area $A$, and uniform solid volume fraction $\nu=\nu_0+\nu_1(t)$, shown in fig.\ \ref{HomPhaseTrans}, as a spring, with a mass $M$ of that portion of granular material. The restoring force is the pressure difference across the flow, $p=p_0+p_1(\nu_1)$. Terms with subscript 0 are associated with the steady state, while those with subscript 1 are perturbations. 

At equilibrium, the gravitational force is balanced by the steady state pressure, so the vertical force associated with a fluctuation in pressure, $p_1 A$, causes a fluctuation, $H_1$, in total flow depth. Because the mass, $M=\rho_s \nu AH$, remains constant and $H_1(t)$ is assumed so small that its products can be neglected, $\ddot{H_1}=-\ddot{\nu_1}H_0/\nu_0$. The fluctuation in volume fraction, $\nu_1$, is then described by
\begin{equation} \label{HomPhaseTransODE}
	\ddot{\nu_1} = -\frac{1}{\rho_s H_0^2}\, p_1\left(\nu_1\right) \, .
\end{equation}

\section{Pressure-Volume Fraction Relationship \label{p-nu}}
We model the dense, collisional gas using an extension of the kinetic theory outlined by Jenkins and Berzi \cite{Jenkins2}, which involves an additional length scale in the rate of collisional dissipation. This length scale is associated with chains or clusters of particles that experience multiple and/or repeated collisions. The length is determined in a balance between the ordering influence of the flow and the randomizing influence of the collisions. In simple shearing flows, the determination of the temperature in a homogeneous form of the energy balance results in an expression for the ratio of shear to normal stress that is the same as that determined by GDR Midi \cite{Midi} for values of their inertial parameter appropriate to collisional flows. In the extended kinetic theory, only the rate of collisional dissipation is different from the constitutive relations for situations in which there are instantaneous, binary collisions. Consequently, we relate the pressure to the volume fraction and the temperature as in standard kinetic theory, but introduce a form of the volume fraction dependence that links disordered and ordered collisional states through a first order phase transition. 

\subsection{Weakly Bumpy Base: Uninterrupted Phase Transition}
A weakly bumpy base promotes slip at the boundary and a resulting increase in collision frequency within dense flows. As a consequence, particle chains are disrupted, while layering of particles is facilitated.

Kinetic theory for a dense, random, collisional flow provides a monotonic increasing relationship between normalized pressure $p/\rho_s T$ and solid volume fraction $\nu$. At sufficiently high volume fractions, a first-order phase transition to an ordered collisional flow is possible \cite{alder,rintoul}. This phase transition is modeled using a cubic curve to join the random and ordered phases in the equation of state for the pressure. Using a cubic curve to model phase change is not a new idea; the long-range attraction and short-range repulsion between molecules of a classical Van der Waals gas may be described with such a curve. However, using a phase change to drive oscillations has not previously been done. 

\begin{figure}			
	\begin{center}
	\includegraphics[width=8cm]{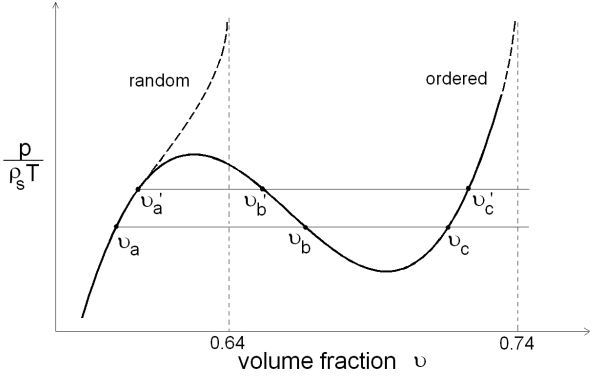}
	\caption{Cubic relationship between $p/\rho_s T$ and solid volume fraction $\nu=\nu_0+\nu_1$, linking a random collisional regime to an ordered collisional regime as $\nu$ increases. The horizontal line at the current steady state value $p_0/\rho_s T_0$ determines three ``equilibrium'' states. Placing the origin at one of these gives $p_1/\rho_s T_0$ versus $\nu_1$. At low inclination $\nu_0=\nu_a$. As inclination decreases and $\nu_0$ increases to $\nu_a'$, $\nu_b$ decreases to $\nu_b'$ and $\nu_c$ increases to $\nu_c'$. \label{CubicCurve1}}
	\end{center}
\end{figure}

The steady state pressure, $p_0\approx\left(H-y\right)\rho_s\nu_0 g\cos\phi$, and the granular temperature $T_0$ are given by Jenkins and Berzi \cite{Jenkins2} in their extension of the kinetic theory for dense inclined flows. A depth-averaged temperature is used for this homogeneous system:
\begin{equation}	\label{T0}
\bar{T_0}=\left[\frac{25\pi^{3/2}}{192}\frac{\hat{c}}{J^2}\right]^3\frac{\left(1+e\right)^5}{\left(1-e\right)^3}\tan^9\phi\frac{Hg\cos\phi}{4} \:,
\end{equation}
where $\hat{c}$=0.50 is a fitted parameter, and
\begin{equation}
J=\frac{1+e}{2}+\frac{\pi}{4}\left[\frac{\left(3e-1\right)\left(1+e\right)^2}{24-\left(1-e\right)\left(11-e\right)}\right] 
\end{equation}
is a function of the effective coefficient of restitution $e$=0.60, as calculated by Jenkins and Berzi \cite{Jenkins2} for inelastic interactions with friction.

The cubic curve is determined from the values of the outermost equilibrium points, $\nu_a$ and $\nu_c$, and the slopes at these points. One of the equilibria is then taken as the origin, so  the corresponding vertical and horizontal axes are $p_1(\nu_1)/\rho_s \bar{T_0}$ and $\nu_1$, respectively. For $p_1/\rho_s \bar{T_0} = C_1\left(\nu_1-\nu_a\right)\left(\nu_1-\nu_b\right)\left(\nu_1-\nu_c\right)$, Eq.\ (\ref{HomPhaseTransODE}) becomes
\begin{equation}	\label{Cubic-HomODE}
	\ddot{\nu_1} = -\frac{T_0 C_1}{H_0^2} \left(\nu_1-\nu_a\right)\left(\nu_1-\nu_b\right)\left(\nu_1-\nu_c\right) \:,
\end{equation}
where the dimensionless coefficient $C_1$ determines the amplitude of the curve.
 
The steady state volume fraction $\nu_0$ increases as the inclination angle $\phi$ decreases. At a high inclination angle, when the system is in a random collisional state, the relatively low volume fraction $\nu_0$ at the origin is at the stable equilibrium point $\nu_a$. Small perturbations from the origin at $\nu_a$ result in small amplitude oscillations within the random collisional regime. As inclination decreases and $\nu_0$ increases, the origin moves upwards along the curve. Correspondingly, $\nu_b$ decreases and $\nu_c$ increases. 

When $\nu_0$ increases to the extent that the normalized pressure moves past the local maxima, the steady state is then the unstable, or repulsive, equilibrium point $\nu_b$ and the system is in a mixed state of random and ordered colliding particles. Here, small decreases from the origin result in large amplitude oscillation, because an initial decrease of solid volume fraction results in an excursion that extends up to and beyond the stable equilibrium point $\nu_a$. The system oscillates between a purely random collisional state and a mixed random and ordered collisional state. Likewise, an initial increase in volume fraction from $\nu_b$ would be attracted to $\nu_c$ in the ordered state and the system would then oscillate between a mixed random and ordered collisional state and an ordered collisional state.

\begin{figure}
	\begin{center}
	\includegraphics[width=8cm]{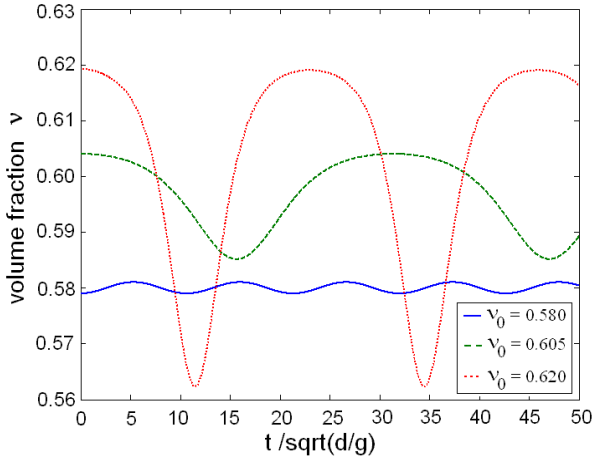}
	\caption{(Color online) Excursions in $\nu$ corresponding to perturbations about steady state $\nu_0$=0.580 (solid line), 0.605 (dashed line) and 0.620 (dotted line), for a flow of 40d depth (d=0.4mm) over a smooth base (no force chains are formed). Local maxima is at 0.598. \label{phasetrans-excursionsVSnue}}
	\end{center}
\end{figure} 

Figure \ref{phasetrans-excursionsVSnue} shows the predicted fluctuations in volume fraction between the random and ordered states for three steady states corresponding to decreasing inclination angle. This behaviour is similar to what Silbert, et al.\ \cite{silbert2} see in inclined flows over a weakly bumpy, frictional base. That is, large oscillations between disordered and ordered states are present past a certain value of $\nu$; while below it, oscillations are small. The situation is different for a bumpier base. A strongly bumpy base facilitates the formation of particle chains that can extend to the base, as seen in fig.\ \ref{patrick-forcechains1}, and their presence interrupts the phase transition. We consider the addition of particles chains in the following section.

\subsection{Strongly Bumpy Base: Interrupted Phase Transition}
A strongly bumpy base reduces slip, collisions, and layering at the boundary, and permits particle chains present within the dense flow to extend to it.

\begin{figure}
	\begin{center}
	\includegraphics[height=4cm]{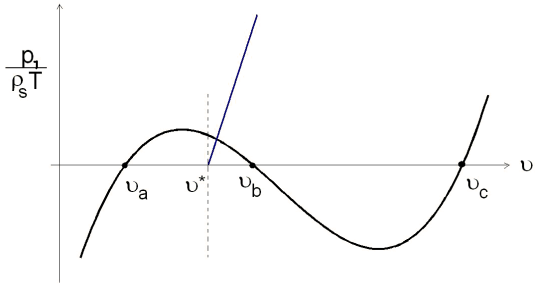}	
	\caption{Relationship between $p_{1}/\rho_s T$ and $\nu$, with the cubic curve representing the collisional regime and the inclined line representing the contact regime. The contribution to pressure arising from particle chains is considered only when $\nu\geq\nu^{\ast}$. \label{2BranchPhaseTrans1b}}
	\end{center}
\end{figure}

The formation of particle chains that span the flow prevents the system from reaching the ordered state. The resistance of these chains is represented by a straight line of larger slope, added to the cubic curve, as shown in fig.\ \ref{2BranchPhaseTrans1b}. The slope of this line is directly proportional to particle stiffness and describes the contribution to the pressure during compression and extension of the particle chains. When $\nu\geq\nu^{\ast}$ particle chains begin to form, so the total normalized pressure is the sum of the contributions from collisions and the force chains.

For Hertzian contact between two identical spheres, the component $P$ of the contact force along the line of centers is related to the corresponding contact displacement by \cite{Love} 
\begin{equation}\label{ContactForce}
\text{P}=M\left(\frac{6\delta}{d}\right)^{3/2} \,,
\end{equation}
where 
\begin{equation}
	M \equiv \frac{2}{9\sqrt{3}}\frac{Gd^2}{1-\psi} \,,
\end{equation}
$G$ is the shear modulus, $\psi$ is Poisson's ratio and $d$ is the particle diameter. Consequently, the increment d$P$ of contact force and the corresponding increment d$\delta$ in contact displacement are related by
\begin{equation}	\label{ContactStressIncrement}
	\text{d}P = \text{K}\text{d}\delta \,,
\end{equation}
where
\begin{equation}
	K\equiv \frac{9M}{d}\left(\frac{6\delta_0}{d}\right)^{1/2} 
\end{equation}
is the contact stiffness and $\delta_0$ is the contact displacement associated with the pressure $p_0$. In such an isotropic compression \cite{Jenkins-Strack}, the contact displacement is related to the volume strain $\Delta_0$ by $\delta_0=\Delta_0 d/6$, and the volume strain is related to the pressure through
\begin{equation}
	\Delta_0 = \left(\frac{\pi d^2}{k\nu_0 M}\, p_0\right)^{2/3} \,,
\end{equation}
where $k$ is the average number of contacts per particle. 

The particles that transmit forces at an angle of $45^{\circ}$ across an area element d$A$ reside in a parallelepiped of volume $\mathbf{k}\cdot\mathbf{N}d\:\left(\text{d}A\right) = d\left(\text{d}A\right)/ \sqrt{2}$, where $\mathbf{k}$ is a unit vector along the line of centers of a contacting pair of particles and $\mathbf{N}$ is a unit normal to the area element. We assume that there are $n$ particles per unit volume, where $n=6\nu_0/(\pi d^3)$. Consequently, the relationship between the stress increment d$\sigma$ in this direction and the contact force increment d$P$ along the line of contact is $\text{d}\sigma = \left(nd/\sqrt{2}\right)\text{d}P$. We take d$\delta=-\left(\nu_1-\nu^{\ast}\right)d$, so 
\begin{equation} \label{AnisotropicStress1}
	\text{d}\sigma = -\frac{3\sqrt{2}\nu_0 K}{\pi d}\left(\nu_1-\nu^{\ast}\right) \ .
\end{equation}
	
Because the particles forming a chain continue to experience collisions, the total increment in pressure fluctation is given by
\begin{eqnarray}	\label{Collision+Contact1}
	p_1 & = & \rho_S T_0 C_1\left(\nu_1-\nu_a\right)\left(\nu_1-\nu_b\right)\left(\nu_1-\nu_c\right)		\nonumber	\\
	& & + \frac{3\sqrt{2}\nu_0 K}{\pi d}\left(\nu_1-\nu^{\ast}\right)		\ .
\end{eqnarray}

We note that particle stiffness affects the system, even if it is purely collisional. More compliant particles overlap more during collisions and, consequently, more effectively fill the space. Hence, the effective solid volume fraction is reduced with higher stiffness. We incorporate this by contracting the axis of volume fraction relative to the left-most equilibrium point when the particle stiffness is increased. This influences both the collisional and contact branches.

\begin{figure}
	\begin{center}
	\includegraphics[width=8cm]{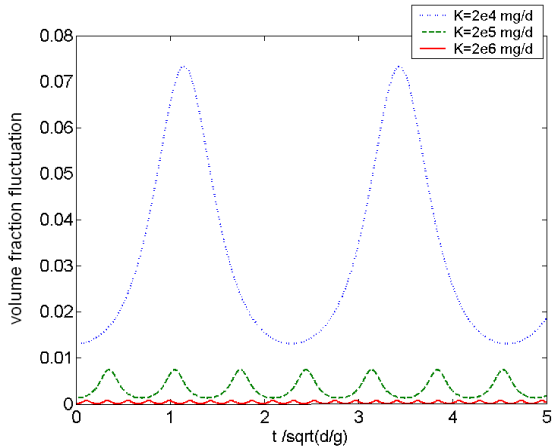}
	\caption{(Color online) Oscillations in packing fraction corresponding to small perturbations about the unstable force-equilibrium point $\nu_b$. Here $\nu^{\ast}$ is taken to be 0.05 of the distance from the local maxima to $\nu_b$. The stiffnesses are $K$=2$\times 10^{4}$ mg/d (dotted line), 2$\times 10^{5}$ mg/d (dashed line) and 2$\times 10^{6}$ mg/d (solid line); equilibrium points are $\nu$=0.566, 0.600 and 0.667; and the initial conditions are $\nu_1=0.75(\nu_b-\nu^{\ast})$, and $\dot{\nu_1}=0$. \label{phasetrans-nuVSk}}
	\end{center}
\end{figure}

We write Eqs.\ (\ref{Cubic-HomODE}) and (\ref{Collision+Contact1}) in terms of dimensionless quantities, and obtain numerical solutions for several values of the stiffness. The stiffness $K$ is specified in units of $mg/d$, where $m$ is the mass of a single particle, for direct comparison with the results of Richard, et al.\ \cite{patrick}. The packing fraction variation with time for a flow of 40d depth and inclination 17.5$^\circ$ is plotted in fig.\ \ref{phasetrans-nuVSk}. Figure \ref{phasetrans-nuVSk} shows that increasing particle stiffness results in a higher oscillation frequency and a smaller oscillation amplitude. The frequencies are close to those measured by Richard, et al.\ \cite{patrick} in numerical simulations, and the decrease in amplitude with stiffness is similar to what they see.

\section{Conclusion}
We have provided a relatively simple explanation of the oscillations seen by Silbert \cite{silbert1,silbert2} and Richard, et al.\ \cite{patrick} in their numerical simulations. The model that we introduce provides an answer to the question of why oscillations of significant amplitude are initiated, and it improves upon that employed by Richard, et al.\ \cite{patrick} in incorporating the nonlinearity associated with the phase change in addition to that associated with the contact force in the chain. It has the capability of reproducing the oscillations and their behavior with changes in parameters observed in the simulations.

Richard, et al.\ \cite{patrick} indicate that their simulations can reproduce the frequencies of audible sound if the stiffness employed is much less than that which result from Hertzian contact using the properties of the bulk material. This suggests that more compliant surface layers on the grains may be playing a role. However, our proposed model implies that all granular aggregates that are composed of nearly identical, round grains are capable of such oscillations. But not all such sands are capable of producing sound, and it may be that the oscillations must be coupled with a wave guide \cite{Caltech} or be in resonance with elastic waves in the bed \cite{Andreotti1} in order to do so. 

We have introduced a model that explains the occurrence of oscillations seen in numerical simulations of idealized systems of spheres. The extension of the model to describe `booming' is speculative. However, it is often said that it is necessary that 'booming' sands be ``well-rounded" and ``mono-disperse" (e.g., \cite{Nori,NetSource1}), and a range of particle sizes suppresses the oscillations seen in the numerical simulations \cite{patrick}. This, at least, is consistent with the mechanism of a phase change between identical spheres on which the model is based.

{\footnotesize{\noindent We are grateful to Mark Shattuck and Sean McNamara for stimulating discussions on, respectively, phase transformations and waves in granular systems. P.\ Richard acknowledges the Department of Theoretical and Applied Mechanics of Cornell University for its hospitality, and the CNRS (PICS France-U.S.) and the ANR (project STABINGRAM No. 2010-BLAN-0927-01) for their support.}}

\bibliographystyle{epj}
\bibliography{Ref-phasetrans-hom}
\end{document}